\documentclass[prl,aps,amssym,nofootinbib,floatfix]{revtex4} 
\usepackage{natbib,epsfig}

\newcommand{\be}{\begin{equation}}
\newcommand{\ee}{\end{equation}}
\newcommand{\bea}{\begin{eqnarray}}
\newcommand{\eea}{\end{eqnarray}}
\begin{document}
\title{Luttinger liquid parameter for the spin-1 Heisenberg 
chain in a magnetic field}
\author{Ian Affleck}
\affiliation{Department of Physics and Astronomy, University of British
Columbia, Vancouver, B.C., Canada, V6T 1Z1}
\date{\today}
\begin{abstract}
A recently derived general formula and older numerical results 
are combined to deduce the behavior of the transverse correlation 
exponent for the spin-1 Heisenberg antiferromagnetic chain in 
an applied magnetic field: $\eta \approx  1/2-(2.0)m+O(m^2)$, where 
$m$ is the magnetization per site.  A comparison with 
the O(3) non-linear $\sigma$-model is also made.
\end{abstract}
\maketitle

We consider integer spin Heisenberg antiferromagnetic chains in a magnetic 
field with Hamiltonian:
\be H=\sum_i[J\vec S_i\cdot \vec S_{i+1}-hS^z_j].\ee
In zero field, this model has  a singlet 
groundstate and a gap, $\Delta$ to the lowest excited state 
which is a spin-triplet of magnons. $\Delta \approx .41J$ for the 
$S=1$ Heisenberg model. The 
gap closes for $h>\Delta$ at which field the $M=1$ magnons 
condense \cite{Affleck1}. ($M$ is the magnetization.) 
 The magnetization per site, $m$, is given by the density 
of these bosons.  In this condensed phase the correlation functions 
exhibit power-law decay:
\bea <S^x_jS^x_0>&=&<S^y_jS^y_0>\propto {(-1)^j\over |j|^{\eta}},
\nonumber \\
<S^z_jS^z_0> &\approx& m^2+{1\over \eta (2\pi j)^2}+{\hbox{const}
\times \cos (2\pi mj)\over |j|^{1/\eta}}.\label{expon}\eea
The exponent, $\eta$ varies continuously with magnetization 
but was argued, based on earlier results of Haldane \cite{Haldane},
 to behave as:
\be \eta =1/2 - O(m).\ee
The validity of this dilute boson picture of the ground state 
was confirmed by density matrix renormalization group calculations 
in \cite{Sorensen,Lou}.  In \cite{Lou} it was shown that 
the interactions between pairs of dilute magnons could be 
conveniently parameterized by a scattering length, $a$.
This was defined in terms of the phase shift, $\delta (k)$ for 
a pair of bosons in vacuum, in the small $k$ limit.  We write 
the 2-particle symmetric wave-function as:
\be \psi (x)\to \sin [k|x|+\delta (k)],\ \  (|x|\to \infty ),\ee
where $x$ is the separation of the 2 particles. 
As $k\to 0$, the phase shift behaves as:
\be \delta (k)\to -ak,\ee
defining the scattering length $a$. (Note that the 
zero interaction case corresponds to $\delta (k)=\pi /2$ for all 
$k$;  $k\to 0$ and the limit of vanishing interaction strength don't 
commute.)  In some cases $a$ 
can be determined exactly, as for general Bose-Hubbard models 
\cite{Affleck2}. 
In cases like the Heisenberg model where 
the bosons are collective excitations $a$, can be determined numerically 
from the finite size corrections to the lowest 2-magnon energy (with 
either open or periodic boundary conditions). 
With periodic boundary conditions, the lowest 2-magnon excitation energy,
(i.e. the energy of the lowest state with $M=2$) 
to $O(1/L^3)$ (where $L$ is the system size) is:
\be E_2-E_0 \approx 2\sqrt{\Delta^2+[\pi v/(L-2a)]^2}\approx 
2\Delta +{1\over \Delta}\left(v\pi \over L\right)^2+\left(4a\over \Delta L
\right)\left(v\pi \over L\right)^2+O(1/L^4).\ee
[The spin-wave velocity 
 $v\approx 2.5J$ for the $S=1$ Heisenberg model.]
A corresponding formula was also derived in \cite{Lou} for 
the case of open boundary conditions.

For the S=1 chain $a$ was determined by this method to be \cite{Lou}:
\be a\approx -2.0,\ee
in units of lattice sites, corresponding to about $-\xi /3$ where 
$\xi \approx 6.1$ is the correlation length in zero field. 
[No systematic attempt was made to estimate the error in $a$, 
which was determined from system sizes $L\leq 100$.  However, two 
different estimates, from DMRG data with 
 periodic and open boundary conditions, differed by $7$\%.] 
$a$ is the basic interaction parameter which determines 
all low density properties of the system.  In particular, 
it was shown \cite{Lou} that it controls the leading correction, at low 
density, to the free fermion result for the magnetization 
density versus field:
\be m\approx {1\over \pi v}\left[\sqrt{2\Delta (h-\Delta)}-{8\Delta 
a\over 3\pi v}(h-\Delta)\right].\label{m}\ee  
This formula was shown \cite{Lou} 
to be in good agreement with numerical data 
for the above value of $a$. 

In \cite{Affleck2} it was shown that $a$ also determines 
the leading correction to the correlation exponents, at 
low densities. There we expressed the exponent $\eta$ in 
terms of a ``Luttinger liquid parameter'' $g$, 
\be 1/\eta \equiv 2g,\label{gdef}\ee
and derived the result:
\be g=1-2am +O(m^2).\label{g}\ee

Combining this exact low density formula for the 
Luttinger liquid parameter $g$ with 
the numerically determined scattering length leads to the 
main result of this brief report:
\be g\approx 1+4.0m + O(m^2),\label{a}\ee
for the nearest neighbor $S=1$ Heisenberg antiferromagnet. 
The corresponding exponents, Eq. (\ref{expon}), might 
be measurable in neutron scattering (or other) experiments. 
This scattering length is {\it not} 
universal, and is modified by crystal field interactions, 
longer range Heisenberg exchange interactions, etc. 
On the other hand, our finite size method for 
determining $a$ can be straightforwardly extended to 
other Hamiltonians. 

This method of determining $g$ at low densities should 
be much more accurate numerically than conventional 
techniques based on direct measurements 
in the finite $m$ system.  These conventional methods 
require 
\be 1\ll M\ll L/\xi ,\label{regime}\ee
 where $L$ is the system size.
 On the 
other hand, $a$ can be determined reliably from 
studies of the $M=2$ sector (with periodic boundary 
conditions, or $M=3$ with open boundary conditions) as long 
as $L$ is large compared to the range of the inter-magnon
interaction, which is of order $\xi$. 
(More precisely, we require $L\gg 2\xi$ for periodic 
boundary conditions or $L\gg 3\xi$ for open boundary conditions.) 
 Numerical results \cite{Fath}
for the Luttinger liquid parameter 
versus magnetization density, $g(m)$, determined from 
the ``Friedel oscillation'' exponent using
Density Matrix Renormalization Group 
 for a chain of length $120$ and $240$ are shown in Fig. (1), 
compared to Eq. (\ref{a}).  The agreement at $m\approx .05$ is 
fair although it seems clear that data for still longer 
chains would be neccessary to test Eq. (\ref{a}).

The low energy theory for the 
large-S limit of integer-spin Heisenberg antiferromagnets is 
given by the O(3) non-linear $\sigma$-model (NL$\sigma$M) 
(with zero topological 
term) with Lagrangian density:
\be {\cal L}={1\over 2\lambda }[(1/v)(\partial_t \vec \phi )^2
-v(\partial_x \vec \phi )^2],\ee
with the constraint $\vec \phi^2=1$. The exact 2-particle 
S-matrix for this (integrable) model \cite{Zamolodchikov} 
 determines \cite{Lou} the scattering 
length $a=-2\xi /\pi$, roughly twice as big as for the 
$S=1$ Heisenberg model.  This discrepancy is presumably 
due to the fact that $S=1$ does not satisfy $S\gg 1$.
Konik and Fendley \cite{Konik} determined $g$ for all magnetic fields 
in the NL$\sigma$M, obtaining in particular a formula 
 for  fields slightly above the critical field $\Delta$:
\be g\to 1+{2^{5/2}\over \pi^2}\sqrt{{h\over \Delta}-1}+O(h-\Delta).\ee
Using the result, $a=-2\xi /\pi$, deduced in \cite{Lou} 
from the exact S-matrix of \cite{Zamolodchikov}, 
together with Eq. (\ref{m}), and $\xi = v/\Delta$ 
(which is an exact consequence of Lorentz invariance), we see
that this result is implied by the general formula 
for $g$, in Eq. (\ref{g}). 

Finally, we remark that this method of determining the 
Luttinger liquid parameter at low densities, using 
Eq. (\ref{g}) and determing the scattering length 
 from studying  the 2-boson sector,
might also be useful for various other models including 
spin ladders in a magnetic field 
 and  bosons in one-dimensional optical lattices, 
close to the ``Tonks gas'' limit. 

I would like to thank  W. Hofstetter, D.R. Nelson, S. Qin and 
U. Schollw\"ock for their collaboration
 and G. F\'ath and E. S\o rensen for helpful communications. This research 
was supported by the Canadian Institute for Advanced Research 
and NSERC of Canada. 

\begin{figure}[ht]
\noindent
\epsfxsize=0.45\textwidth
\epsfbox{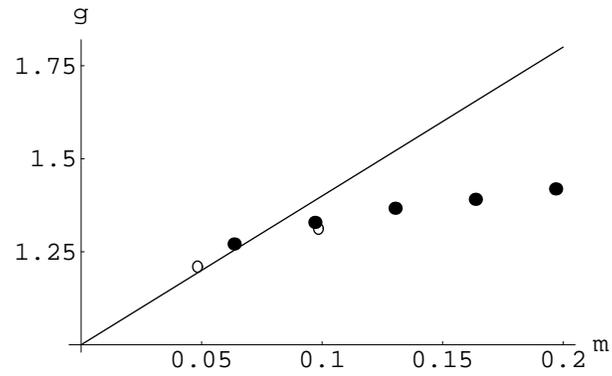}
\caption{Luttinger liquid parameter, $g$, versus magnetization per site, $m$. 
Solid circles are numerical data for a chain of length 120 
and open circles for length 240  
from \cite{Fath}; line is Eq. (\ref{a}).}
\label{fig:beta}

\end{figure}

\end{document}